\documentclass[twocolumn]{jpsj2} 
\title{Appearance of New Energy Gap and Periodic Local Density-of-States Modulation in Bi$_{2}$Sr$_{1.6}$La$_{0.4}$CuO$_{6+\delta}$}

\author{\textsc{Tadashi Machida}, \textsc{Yusuke Kamijo}, \textsc{Keiji Harada}, \textsc{Tetsurou Noguchi}, \textsc{Ryo Saito}, \\ \textsc{Takuya Kato} and \textsc{Hideaki Sakata}\thanks{E-mail address: sakata@rs.kagu.tus.ac.jp}}

\inst{Department of Physics, Tokyo University of Science, \\
1-3 Kagurazaka, Shinjyuku-ku, Tokyo, 1628601\\}

\abst{ The spatial variation of the local density of states in optimally doped Bi$_{2}$Sr$_{1.6}$La$_{0.4}$CuO$_{6+\delta}$ (superconducting transition temperature is 34 K) is studied by scanning tunneling spectroscopy at 4.2 K in zero magnetic field.
Two-dimensional density-of-states modulation aligned with the Cu-O-Cu bond with a periodicity of about five lattice constants is observed. 
It is found that this modulation accompanies the appearance of a new energy gap of approximately 10 meV, whose gap edge peak is spatially modulated in intensity. This gap coexists with the superconducting gap, the value of which ranges from 10 meV to 60 meV.
}

\kword{STM, STS, cuprate, Bi$_{2}$Sr$_{1.6}$La$_{0.4}$CuO$_{6+\delta}$, local density-of-states modulation }

\begin{document}
\maketitle

The addition of a relatively small number of carriers into a Mott insulator realizes $\it{d}$-wave high-temperature superconductivity in cuprates.
The doped cuprates show unconventional electronic states.
This unconventional nature has been investigated extensively in both momentum and real space. 
Angle-resolved photoemission spectroscopy (ARPES) revealed that states in different parts of momentum space, i.e., nodal and antinodal regions, show contrasting features, such as different temperature and doping dependences~\cite{Norm}.
Furthermore, the line shape of photoemission spectra indicates dichotomy; a broad line shape in the antinodal region changes abruptly to a sharp one near the nodal region~\cite{Zhou}.
In addition, the superconducting gap structure along the Fermi surface considerably deviates from the prismatic $\it{d}$-wave functional form~\cite{Harr,Ronn}.

On the other hand, scanning tunneling spectroscopy (STS), which measures integrated spectral weight in momentum space, revealed spatial evolution of the electronic states.
The superconducting gap varies on a length scale of about 5 nm and the tunneling spectrum changes distinctly in its shape~\cite{Pan,Lang}.
It is claimed that coherence peaked spectra are observed in the region to be occupied by a canonical $\it{d}$-wave superconductor ($\it{d}$SC), while the spectra without clear coherence peaks, attributed to the destruction of antinodal coherence peaks, is characteristic of the electronic phase at the zero temperature limit of the pseudogap (ZTPG) region~\cite{Mce1}.
Furthermore, an anomalous local density-of-states (LDOS) modulation with a periodicity of about four lattice constants has been observed in Bi$_{2}$Sr$_{2}$CaCu$_{2}$O$_{8+\delta}$ (Bi2212) and Ca$_{2-x}$Na$_{x}$CuO$_{2}$Cl$_{2}$ (Na-CCOC)~\cite{Mce1,Hof1,Hof2,Hana,Howa,Fang,Momo}.
This modulation was first observed by Hoffman $\it{et}$ $\it{al.}$ in a magnetic field around vortex cores~\cite{Hof1}.
In zero magnetic fields, both energy-dispersive and nondispersive LDOS modulations were reported. 
The former, observed at low energy, is attributed to the interference of the quasi particles at the Fermi arc regions~\cite{Hof2,Mce2}.
The dispersion of the quasi particles obtained in the STS experiments and the ARPES results are in excellent agreement. 
The origin of the latter, however, is still controversial.
Since the nondispersive LDOS modulation has been observed in pseudogap states above $\it{T_c}$ and in the underdoped non-superconducting region~\cite{Vers,Hana}, it is claimed that this modulation is associated with the pseudogap.
Fermi surface nesting, pinned stripes and more novel electronic phases have also been proposed to explain the modulation~\cite{Shen,Howa}.

In the above-cited experiments, however, the observed intensity of the LDOS modulations is so weak that it is difficult to investigate their energetic structure.
Here, we report on STS measurements in Bi$_{2}$Sr$_{1.6}$La$_{0.4}$CuO$_{6+\delta}$ (Bi2201-La) with $ \textit{T}_{c} $ of 34 K in zero magnetic field.
LDOS modulation two-dimensionally aligned with the Cu-O-Cu bond with a period of about five lattice constants was observed.
We found that this LDOS modulation accompanies the appearance of a new energy gap of approximately 10 meV whose gap edge peak height is spatially modulated in intensity. This gap coexists with a superconducting gap whose value ranges from 10 meV to 60 meV.

\begin{figure}[tb]
\begin{center}
\includegraphics[width=8cm]{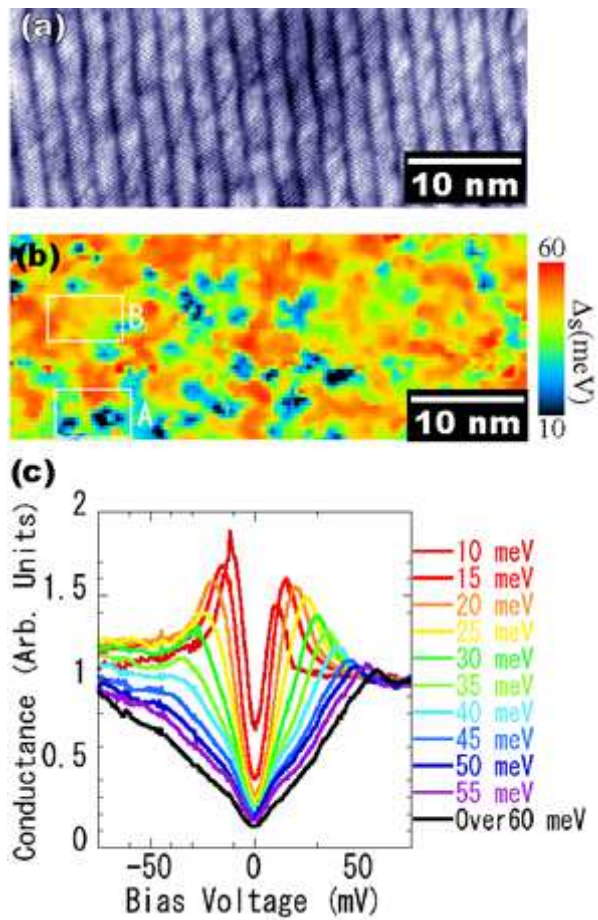}
\end{center}
\caption{(a) Typical topographic image of the cleaved sample surface with atomic resolution (46 nm $\times$ 13 nm). Bias voltage is 250 mV. Superlattice modulation with a period of 2.2 nm is along the $\it{b}$-axis. (b) Gap map for the same field of view as (a). Local gap value $\Delta_s$ from 10 to 60 meV is shown in color scale. The gap value is defined by the energy at the coherence peak above the Fermi energy. (c) The averaged spectra of all regions exhibiting a given local gap value. The spectra are normalized at 80 mV. Kink or shoulder structures are indicated by arrows.}
\label{f1}
\end{figure}

Single-crystalline Bi2201-La was grown by the floating zone method.
The substitution of trivalent La for divalent Sr can reduce the hole concentration.
The concentration of La, verified by inductively coupled plasma mass spectrometry, was 0.4 so that the sample is nearly optimally doped~\cite{Name}.
The transition temperature of 34 K was determined by magnetization measurements.
A laboratory-made LT-STM was used for STS measurements~\cite{Saka}.
Bias voltage was applied to the sample and all measurements in this study were done at 4.2 K.
Tunneling spectra were measured at 128 $\times$ 36 points in a 46 $\times$ 13 nm$^2$ area, thus spatial resolution is 0.36 nm.
We cleaved the samples in situ at 4.2 K. 
We obtained almost the same results in multiple areas on the same sample.
Figure 1(a) is a typical topographic image of the sample surface with atomic resolution.
The topograph shows the incommensurate supermodulation with a period of 2.2 nm along the $\it{b}$-axis.

The spatial variation of the superconducting gap value in Bi2212 and La$_{2-x}$Sr$_{x}$CuO$_{4} has been reported$~\cite{Lang, Kato}.
Bi2201-La showed almost the same variation.
In Fig. 1(b), we show a gap map for the same field of view as in Fig. 1(a).
The gap value, $\Delta_s$, defined by the energy at the coherence peak above the Fermi energy, is distributed from 10 meV to 60 meV, and varies on a length scale of about 3 nm.
The averaged spectrum of all regions exhibiting a given local gap value is shown in Fig. 1(c).
This figure indicates the following features.
(i) Coherence peaks diminish in strength with increasing $\Delta_s$. 
(ii) The spectra having $\Delta_s$ larger than 45 meV show almost no peak at both the positive and negative gap edges.
(iii) A weak kink or shoulder structure, conspicuous when $\Delta_s$ is larger than 25 meV, exists at about $\pm$10 meV, and this energy appears to be constant regardless of $\Delta_s$.
These features are similar to those in Bi2212~\cite{Mce1}.
Thus, the observed electronic disorder is the same entity as that observed in Bi2212.

\begin{figure}[tb]
\begin{center}
\includegraphics[width=7cm]{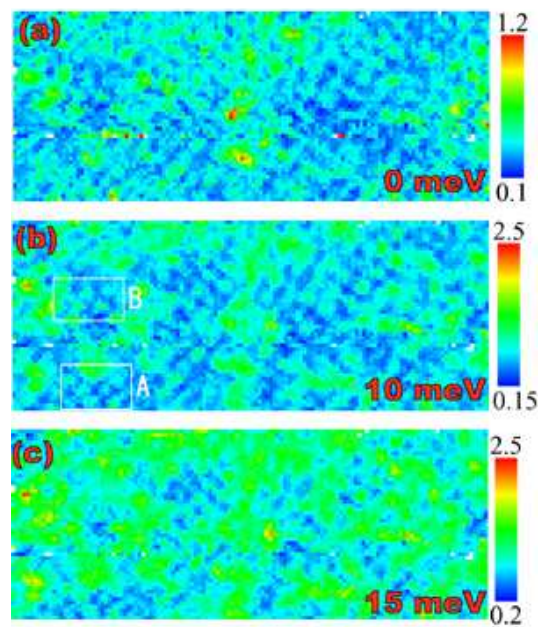}
\end{center}
\caption{Conductance maps at (a) 0 meV, (b) 10 meV and (c) 15 meV for the same field of view as in Fig. 1(b). The maps indicate the normalized conductance (see text). Note a color scales are not identical in the figures.}
\label{f2}
\end{figure}

Figures 2(a)-2(c) show conductance maps for the same field of view as in Fig. 1(a) at 0, 10 and 15 meV.
The conductances for each map are scaled by their respective magnitudes at 80 meV.
Thus, the conductance maps represent N($\it{r},\it{E}$)/N($\it{r}$,80meV), where N($\it{r},\it{E}$) is LDOS at position $\it{r}$ and energy $\it{E}$.
Periodic modulation along the Cu-O-Cu direction can be seen at each map.
The modulation is partly hidden due to the spatial variation of the superconducting gap, particularly at 15 meV.
Although the observed pattern is two dimensional, the lattice constant and the direction show slight local variation.
Fourier analysis of the maps reveals that the averaged period of the modulation is (5.2$\pm$0.2)$\it{a_0}$ (where $\it{a_0}$ is the Cu-O-Cu distance).
Line cuts along the Cu-O-Cu direction of the Fourier-transformed images of the conductance maps are shown in Fig. 3.
This figure shows that the period of modulation is energy-independent within our experimental accuracy.

\begin{figure}[tb]
\begin{center}
\includegraphics[width=6cm]{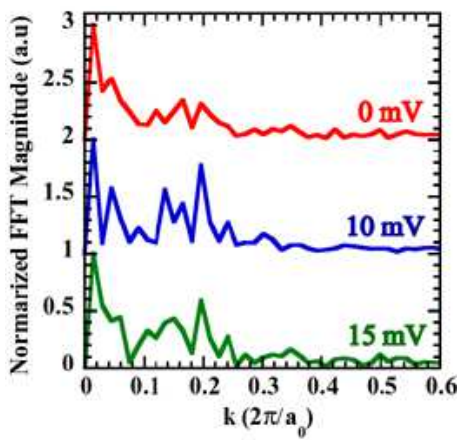}
\end{center}
\caption{Line cuts along the Cu-O-Cu direction of the Fourier transformed images of the conductance maps shown in Fig. 2. $\it{a_0}$ is the Cu-O-Cu distance. Each curve is shifted by 1 vertically.}
\label{f3}
\end{figure}

Figures 4(a) and 4(b) show the evolution of the tunneling spectra in region A, depicted in Figs. 1(b) and 2(b), where $\Delta_s$ is about 50 meV.
At the trough of the modulation, the spectrum shows a pseudogap nature; there is no clear coherence peak at the gap edges (the blue spectrum in Figs. 4(a) and 4(b)).
As the measuring point is moved from the trough to the crest, the in-gap states at about $\pm$ 10mV gradually grow, though $\Delta_s$ remains constant. 
At the crest, the in-gap states become clear peaks (the red or orange spectrum), though the peak below the Fermi energy is less conspicuous.
As a result, a new gap feature appears at the crest.
This new gap coexists with the large energy gap $\Delta_s$.

As shown in Fig. 4, the shape of the tunneling spectra changes spatially with the period of about five lattice constants.
Thus, the observed conductance modulation shown in Fig. 2 represents this modulation of the LDOS. 
This is consistent with the nondispersive nature of the observed conductance modulation, because the shape of the tunneling spectrum changes at a particular wavelength.
We note that the energy of the new gap is almost the same as that of the kink appearing on the tunneling spectra shown in Fig. 1(c). 
Indeed, in Figs. 4(a) and 4(b), the kinks at the trough (the blue spectrum) appear to grow continuously to the gap edge peaks (the red spectrum).
Thus, it is thought that the kinks are indicative of the new gap even at the trough of the modulation.

\begin{figure}[tb]
\begin{center}
\includegraphics[width=7cm]{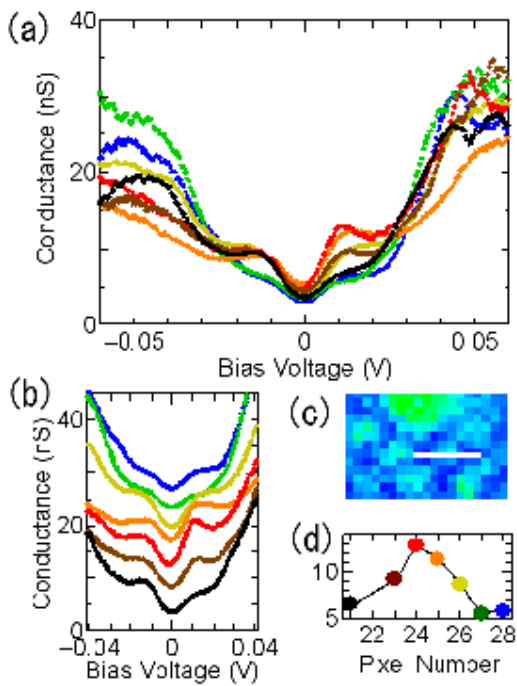}
\end{center}
\caption{(a),(b) Spatial evolution of tunneling spectrum in region A. The spectra were acquired along the white line depicted in (c). In (b), each spectrum is shifted by 5 nS. (c) Conductance map at 10 meV in region A, shown in Figs. 1(b) and 2(b). (d) Conductance at 10 meV as a function of pixel number along the white line in (c). One pixel corresponds to 0.36 nm. The color of the closed circle corresponds to that of the spectra in (a) and (b).}
\label{f4}
\end{figure}

\begin{figure}[tb]
\begin{center}
\includegraphics[width=7cm]{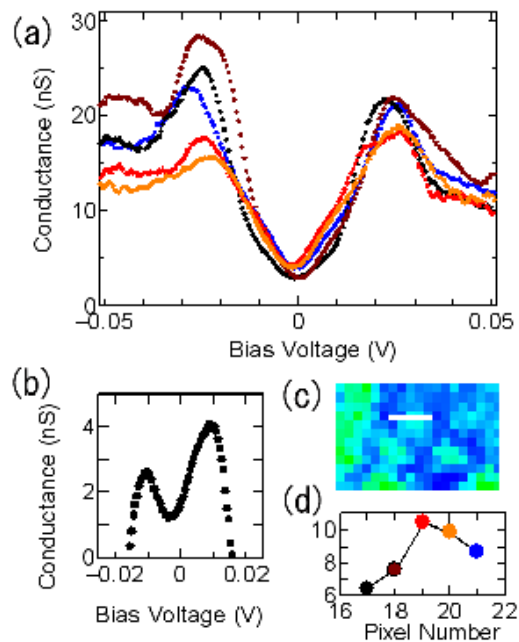}
\end{center}
\caption{(a) Spatial evolution of the tunneling spectrum in region B. The spectra were acquired along the white line depicted in (c). (b) Difference in conductance between spectra at the crest (red) and at the trough (black). (c) Conductance map at 10 meV in region B shown in Figs. 1(b) and 2(b). (d) Conductance at 10 meV as a function of pixel number along the white line in (c). One pixel corresponds to 0.36 nm. The color of the closed circle corresponds to that of the spectrum in (a).}
 \label{f5}
\end{figure}

The LDOS modulation was also observed in region B where the mean $\Delta_s$ is about 25 meV.
Figure 5(a) shows the evolutions of the tunneling spectra in region B.
At the trough of the modulation, the spectrum shows clear coherence peaks.
As the measuring point is moved from the trough to the crest, the height of the coherence peaks decreases.
This change is the same as that reported in Bi2212~\cite{Fang}.
Furthermore, in-gap states between 0 and 15 meV increase at the crest.
Since $\Delta_s$ does not change in this local evolution, the electronic disorder is not the origin of the difference in the tunneling spectra.
Difference between the spectra of the crest and the trough, shown in Fig. 5(b), reveals two peaks at $\pm$ 10 meV.
This indicates that the LDOS modulation in region B also accompanies the appearance of the same energy gap with gap-edge peaks at $\pm$ 10 meV as that in region A.
Almost the same evolution of the tunneling spectrum was observed in the regions having $\Delta_s$ between 25 and 50 meV.
This indicates that the modulation exists regardless og the $\it{d}$SC region or ZTPG region in Bi2201-La.
In all regions, the value of the new gap was almost constant independent of $\Delta_s$.

We note that the value of the new gap is coincident with the lower limit of the variation of $\Delta_s$.
Thus, the clear separation of the two gaps was observed only in the region where $\Delta_s$ is much larger than 10 meV, namely, 40 meV or higher energy.
We have not observed the dispersive LDOS modulation.
This is because the amplitude of the nondispersive LDOS modulation described above is much larger than that of the dispersive one that may have almost the same period.

Obviously, the experimental results obtained raise the following questions: "What is origin of the new energy gap?" and "Why does the gap feature emerge periodically?" 
In the usual sense, an energy gap opens when an electronic ordered state emerges.
Thus, the new gap feature may be attributed to some kind of ordered state coexisting with superconductivity, such as a charge density wave (CDW) state.
In Bi2212 and NCCOC, LDOS modulation is attributed to the opening of the pseudogap in the ZTPG region~\cite{Mce1,Hana}.
However, this is not the case in Bi2201-La, because the new gap is much smaller than the pseudogap and the LDOS modulation is observed regardless of the $\it{d}$SC region or ZTPG region.

We noted that (i) the value of the new gap is almost constant over the surface at our experimental resolution, in spite of the significant variation of $\Delta_s$, and (ii) the value of the new gap is coincident with the lower limit of the variation of the superconducting gap $\Delta_s$.
In a pristine $\it{d_{x^2-y^2}}$-wave superconductor, the superconducting gap varies along the Fermi surface corresponding to the $\it{d}$-wave order parameter of the cos($\it{kx}$)-cos($\it{ky}$) form. 
Since STS measures spectral weight integrated in momentum space, the observed spectrum includes spectral weights from all regions, leading to a V-shaped tunneling spectrum.
In fact, in Bi2212, the $\it{d}$-wave functional form well describes the experimental gap structure~\cite{Din2}, and the tunneling spectrum  shows an almost V-shaped form in the regions with small $\Delta_s$~\cite{Mce1}. 
In Bi2201-La, however, ARPES experiments revealed that the gap variation along the Fermi energy shows considerable deviation from the $\it{d}$-wave functional form; the gap shows flattening near the nodal region and a steep increase near the antinodal region~\cite{Harr}.
Such phenomena have also been reported in NCCOC.
Ronning et al. proposed two different gaps to reconcile the measured angle dependence of the gap; a very small one contributing to the nodal region and a very large one dominating the antinodal region~\cite{Ronn}.
Furthermore, the line shape in ARPES changes abruptly from a broad one in the antinodal region to a sharp one near the nodal region~\cite{Zhou}.
These results suggest that the $\it{d}$-wave superconducting gap is divided into two energy gaps in the tunneling spectrum when the broadening near the antinodal region is much larger than that near the nodal region.
Along this line, it is possible to think that our measurements showed two superconducting energy gaps from different regions in momentum space: $\Delta_s$ attributed to the near-antinodal one and a new small gap contributing to the near-nodal one.
Indeed, electronic states near the nodal region are not affected by disorder in the samples, whereas the near-antinodal states are significantly modified~\cite{Mce1}.

The reason for the periodic nature of the gap-edge peak height is not clear at present.
If the new gap is attributed to the superconducting gap near the nodal region, the modulation of the peak height may be caused by the modulation of the nodal spectral weight or relative change in weight between the nodal and the antinodal region.
This needs further investigation.

In this paper, we presented the relationship between the LDOS modulation and the appearance of a new gap feature.
The results will place strict constraints on theoretical models and provide a crucial key to understanding of the LDOS modulation.

\end{document}